\documentclass[twocolumn,twoside,slac_two]{revtex4}
\usepackage{graphicx}
\usepackage{fancyhdr}
\begin{document}

\title{Suzaku X-Ray Monitoring of Gamma-Ray-Emitting Radio Galaxy, NGC 1275}

%

\author{Ikumi Edahiro, Yasushi Fukazawa, Kenji Kawaguchi, Yasuyuki Tanaka, Ryosuke Itoh, Yuka Kanda, Kensei Shiki}
\affiliation{Hiroshima University/Hiroshima High Energy \&
Optical/Infrared Astrophysics Laboratory, Higashi-Hiroshima, Hiroshima, 739-8526}
%

\begin{abstract}
NGC 1275 is a gamma-ray-emitting radio galaxy at the center of the Perseus cluster.
Its multi-wavelength spectrum is similar to that of blazers,
and thus a jet-origin of gamma-ray emissions is believed.
In the optical and X-ray region, NGC 1275 also shows a bright core,
but their origin has not been understood, since a disk emission is not ruled out.
In fact, NGC 1275 exhibits optical broad emission lines and a X-ray Fe-K line, 
which are typical for Seyfert galaxies.
In our precious studies of NGC 1275 with Suzaku/XIS,
no X-ray time variability was found from 2006 to 2011, 
regardless of moderate gamma-ray variability observed by {it Fermi}-LAT~\cite{Yamazaki}.
We have continued monitoring observations of NGC 1275 with Suzaku/XIS.
In 2013-2014, MeV/GeV gams-ray flux of NGC 1275 gradually increased and reached 
the maximum at the beginning of 2014.
Correlated with this recent gamma-ray activity, we found that X-ray flux also increased, 
and this is the first evidence of X-ray variability of NGC 1275.
Following these results, we discuss the emission component during the
 time variability, but we cannot decide the origin of X-ray variability
 correlating with gamma-ray.
Therefore, for future observation, it is important to observe NGC 1275 by using Fermi gamma-ray,
 XMM-Newton,
 NuStar, ASTRO-H X-ray, CTA TeV gamma-ray and Kanata optical telescope.

\end{abstract}

\maketitle

\thispagestyle{fancy}


\section{Introduction}
Active Galactic Nucleus(AGN) is one of high energy objects.
AGN are thought to be composed of massive black hole, 
emission-line region around the black hole, absorption of torus, and
bidirectional plasma AGN jets.
AGN observed in wide wavelength range from radio to gamma-ray 
and appearance of AGN and the time variation are different for each wavelength.
However, detailed structure of AGN and radiation mechanism of AGN jet
have not been clear with past various observations.
Radio galaxes which is one of types of having AGN are very important to
study AGN jet phenoneta.

\begin{figure}[h]
\includegraphics[width=8cm]{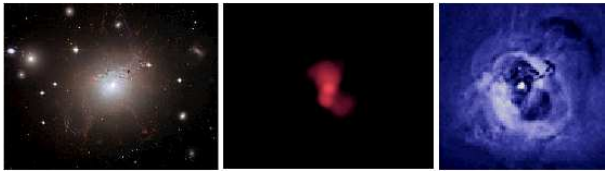}
\caption{The pictures of NGC 1275. (left) Optical band. There are many galaxies.(middle) Radio band. We can see the structure of AGN jet. (right) X-ray band. At the center of this picture, we can see the nucleus.} \label{picture}
\end{figure}

NGC 1275 as shown in Figure~\ref{picture} is an elliptical galaxy, locating at the center of the Perseus
cluster.
A viewing angle of apparent jet is about 30-55 degree~\cite{Walker}.
NGC 1275 is known as an AGN and classified as a radio-loud Seyfert galaxy or a
radio galaxy.
In recent years, Fermi detected GeV gamma-ray emission for the first time and
NGC 1275 is the brightest in gamma-ray among radio galaxies~\cite{Abdo}~\cite{Kataoka}.

NGC 1275 has been observed various wavelength.
Fermi observation shows the time variation of GeV gamma-ray flux with
several months scale
and the gamma-ray flare was also reported~\cite{Kataoka}~\cite{Donato}~\cite{Brown}.
From the above Fermi observations, it is suggested that the gamma-ray
emission does not come from Perseus cluster via the cosmic ray interactions.
Furthermore, TeV gamma-ray was detected with MAGIC~\cite{Ale1} .
Recently, from light curve of Fermi gamma-ray and MAGIC KVA optical
R-band from 2010 to 2011, 
variability correlation between GeV gamma-ray and optical R-band was found~\cite{Ale2}.

Therefore, a radio-loud gamma-ray emitting Seyfert galaxy NGC 1275 is very attracting to
study these structure.
The SED of NGC 1275 nucleus can be explained 
by synchrotron self-Compton(SSC) model~\cite{Abdo}~\cite{Yamazaki}.
The SED of NGC 1275 jet component rely on the radio and gamma-ray band
because of optical and X-ray emission from jet has not been clear.
So, it is important to search X-ray jet flux that could change SSC model parameters.

In the X-ray band,
Einstein detected a point-like source~\cite{Bra}.
However, the past observations could not constrain the X-ray spectrum.
Recently, XMM-Newton and Chandra observed NGC 1275 and could resolve the nucleus emission spacially.
From the results of XMM-Newton observation, 
a photon index of NGC 1275 is 1.65 and a flux is 1.43 $\pm 0.29 \times 10^{-11} \rm{erg/cm^{2}/s}$ 
in 0.5-8 keV band~\cite{Chura}
and the results of Chandra observation, 
a photon index is $1.6 \pm 0.1$ and a flux is $6.1 \times 10^{-12} \rm{erg/cm^{2}/s}$ 
in 0.5-5 keV band~\cite{Balma}.
However, the number of observations are small, simultaneous observations with Fermi are poor,
and Chandra data suffer from pile-up for the nucleus
because of the hosting Perseus cluster is very bright. 
From Swift/BAT spectrum of NGC 1275,
AGN emission is reported to be marginally detected,
but Swift/BAT could not resolve the nucleus spatially.
A photon index is $1.7^{+0.3}_{-0.7}$ and a flux is $1 \times 10^{-11} \rm{erg/cm^{2}/s}$ in 15-55 keV band
~\cite{Aje}.

Therefore, we tried to search nucleus emitting of Perseus cluster by using archival Suzaku/XIS data.
Suzaku/XIS has observed the Perseus Cluster every half year with 40ks.
From Yamazaki+13\cite{Yamazaki}, 
they search variability correlation between Suzaku/XIS X-ray and gamma-ray flare, and 
there seem no correlation between X-ray and gamma-ray 
from 2008 to 2011.
From the results of Suzaku/XIS observation, photon index is 1.6-1.8 and this result is consistent with 
the XMM-Newton resutls.
Recently, from 2013 to 2014, a big GeV gamma-ray flare occurred as shown
in bottom of Figure~\ref{lc} in orange arrow region.
So, we tried to extend the analysis of Suzaku/XIS observation data to stury the variability correlation 
with a big GeV gamma-ray flare.
If we find the variability correlation between X-ray and gamma-ray, it is a key to solve the AGN emission 
mechanism.


\section{Analysis of Suzaku/XIS data of NGC 1275}
\begin{figure}[h]
\includegraphics[width=8cm]{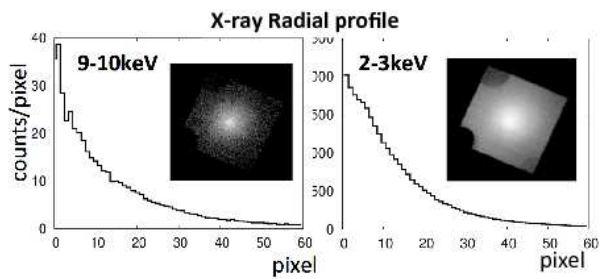}
\caption{X-ray radial profile and image (left) 9-10 keV band (right) 2-3
 keV band} \label{imaging}
\end{figure}

Here is the method of the analysis of Suzaku/XIS data of NGC 1275.
Suzaku PSF cannot resolve NGC 1275 nucleus well.
So, we extracted the AGN emission by imaging spectroscopy.
First, we created images at 2-3, 3-4, 4-5, 5-6, 6-7, 7-8, 8-9,  9-10, 10-12 keV energy band 
and derived radial count profiles.
For example, Figure~\ref{imaging} is X-ray radial profiles and images in 9-10 keV band (left)
and 2-3 keV band (right).
We can find that the hard X-ray AGN emission is seen at the center
region in 9-10 keV band.

\begin{figure}[h]
\includegraphics[width=8cm]{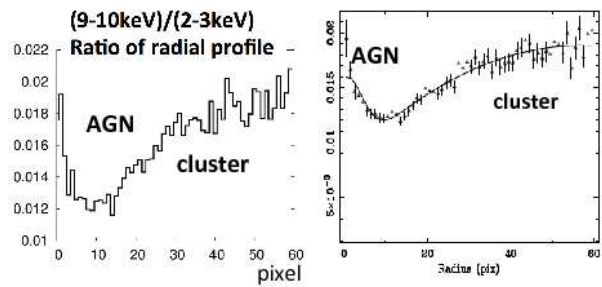}
\caption{(9-10 keV)/(2-3 keV) ratio of radial profile. (left) At the center region, AGN emission is dominant. (right) By fitting with f(x), we obtained the AGN photon counts.} \label{fit}
\end{figure}

\noindent Figure~\ref{fit} is the ratio of these two radial profile.
Hard X-ray is clearly seen at the center region.
We fitted this profile with Gaussian and quadratic function as 
\begin{equation}
f(x) = \frac{2n}{\sqrt{2\pi} \sigma} exp\left[ -\frac{1}{2} \left(\frac{x}{\sigma} \right)^{2} \right]  
 +d\left[ a(x-b)^{2} + c \right] \,\,\,.
\end{equation}
Here, Gaussian express the emission from the AGN at the center region and quadratic function express 
the emission from Perseus cluster at the outside region.
We subtracted the model-cluster component from the data, 
and obtained the AGN photon counts.
We analyzed the data from 2006 to 2014 and derived an X-ray light curve.
This method is not usual, and we confirmed whether the results of using this method are good or not by applying the same way to brazar 3C 454.3.
From comparing with the results of standard way and above way, 
we got the consistent results with Chandra, XMM-Newton and Swift/BAT observations  results
as same as Yamazaki+13~\cite{Yamazaki}. 

\section{Results and discussion} 
\subsection{Suzaku/XIS X-ray Light Curve of NGC 1275}
\begin{figure}[h]
\includegraphics[width=8cm]{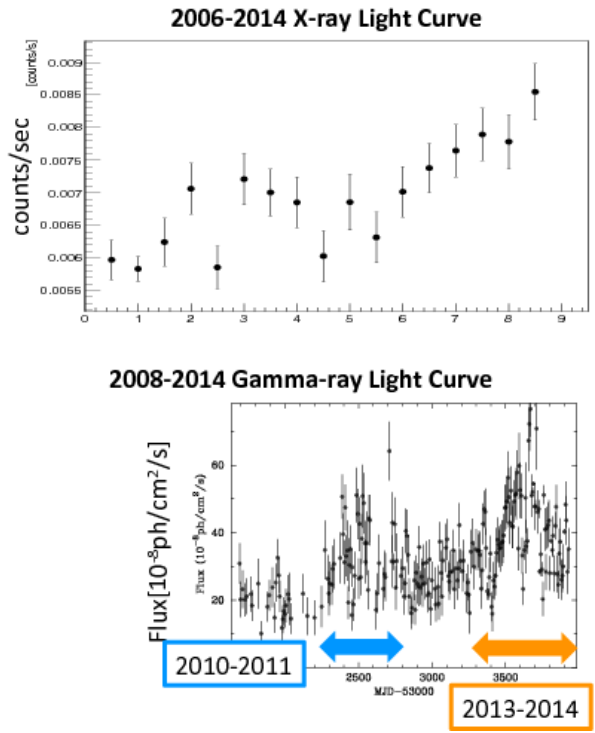}
\caption{Light curve of NGC 1275 (top) 2006-2014 Suzaku X-ray (bottom)
 2008-2014 GeV gamma-ray (archival light curve supplied by GSFC). 
 The bin size of the horizontal axis are collect between top and bottom.} \label{lc}
\end{figure}

The top of Figure~\ref{lc} is Suzaku X-ray light curve from 2006 to 2011.
Compared with Suzaku X-ray and Fermi gamma-ray light curve(as shown in bottom of
Figure~\ref{lc}), we can see brightening of the nucleus in the X-ray band in
2013-2014 in orange arrow region, correlating with GeV gamma-ray flare.
This is the first evidence of X-ray variability of NGC 1275.
X-ray spectrum is consistent with the XMM-Newton results.
However, it is not clear how the X-ray spectrum varied, because of
Suzaku/XIS PSF.
In 2010-2011 at the blue arrow region in bottom of Fig.~\ref{lc}, there are no correlation and this result is consistent with Yamazaki+13~\cite{Yamazaki}.

\subsection{Disccussion of Origin of X-ray Variability}
\begin{figure}[h]
\includegraphics[width=7.5cm]{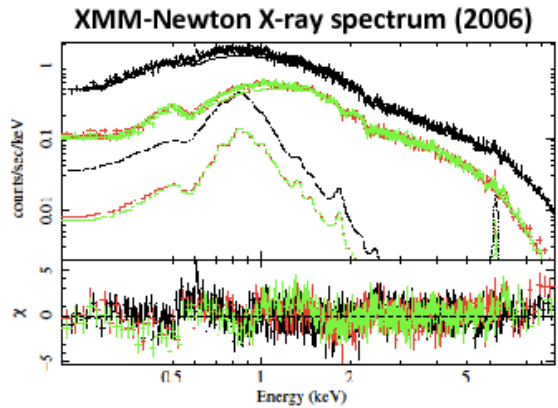}
\caption{XMM-Newron X-ray Spectrum of 2006.} \label{xmmspe}
\end{figure}

We consider what is the origin of X-ray variability.
Figure~\ref{xmmspe} is XMM-Newton spectra of NGC 1275 in 2006.
From Fig.~\ref{xmmspe}, the spectrum is well described by the simple power-law with
a photon index of 1.73 and a Fe-K line of equivalent width of 70 keV.
This is similar to that of Seyfert galaxies.
Because of weak correlation between X-ray and gamma-ray in 2008-2011,
disk/corona emission seems to be dominant in the X-ray band.

On the other hand, what is the origin of X-ray variability correlating
with gamma-ray?
The possible origin is jet emission or disc/corona emission.
If disc/corona emission is the origin of X-ray variability correlating
with gamma-ray, NGC 1275 would become a rare object from which
both disk/corona emission and jet emission from X-ray to gamma-ray band.
We can study the disk/jet connection from the X-ray and gamma-ray
correlation.
Optical lines are reprocess of disk/corona emission, while X-ray traces
the disk/corona emission directly.

\begin{figure}[h]
\includegraphics[width=8.5cm]{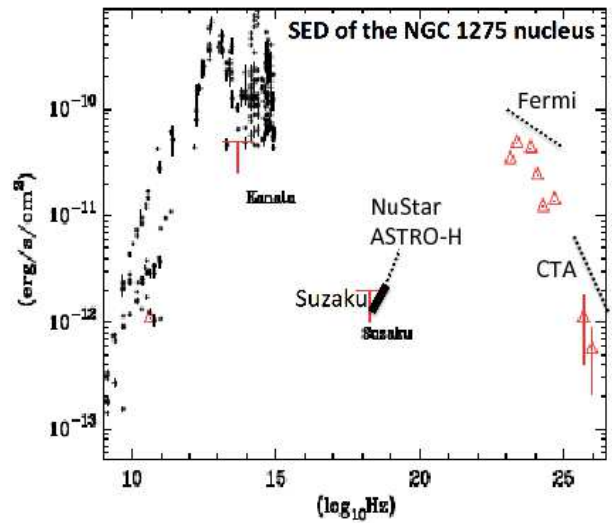}
\caption{SED of the NGC 1275 nucleus~\cite{Yamazaki} and observation region with each satellites.} \label{sed}
\end{figure}

If jet emision is the origin of X-ray variability correlating with gamma-ray, variable X-ray component would be a low energy tail of
inverse Compton.
We can trece a precise SED variability from X-ray to gamma-ray to
constrain the flare mechanism.
In the near future, we can trace the jet flare from X-ray and gamma-ray,
with NuStar, ASTRO-H, Fermi, and CTA as shown in Figure~\ref{sed}.

\begin{figure}[h]
\includegraphics[width=8.5cm]{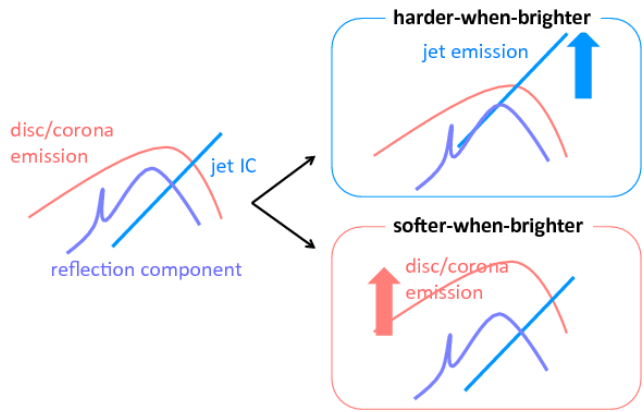}
\caption{Inference of X-ray components(disc/corrona(red), jet(blue), and reflection(purple)) for NGC 1275 and correlation between variable component and energy band.} \label{component}
\end{figure}

We infer that the following X-ray spectral components for NGC 1275,
disk/corona, reflection, and jet, as shown in Figure~\ref{component} (left).
If we could obtain the X-ray spectral variability of
harder-when-brighter as shown in Fig.~\ref{component} upper right, 
the variable component is jet inverse compton.
If softer-when-brighter as shown in Fig.~\ref{component} lower right,
the variable component is disk/corrona emission.
So, future X-ray observations, for example, XMM-Newton, NuStar, ASTRO-H,
are importan.
In addition, it is important to observe NGC 1275 by observed optical
region.

\section{Conclusions}
We analyzed Suzaku/XIS observation data of NGC 1275.
From 2013 to 2014, brightening of the nucleus in the X-ray band was
found, correlating with GeV gamma-ray flare.
This is the first evidence of X-ray variability of NGC 1275.
However, we cannot find what the variability component is, disc/corrona
or jet.
For future prospects, in addition to Fermi observation, it is important
to observe NGC 1275 by using XMM-Newton, NuStar, ASTRO-H.
CTA TeV gamma-ray observation is also important to understand the
gamma-ray flare.
We are also continuing to monitor NGC 1275 by Kanata optical telescope.
By using simultaneous multi wavelength spectra, we want to find the
detail structure of AGN.



\end{document}